\documentclass[slac_one]{revtex4}
\usepackage{graphicx}
\usepackage{fancyhdr}
\newcommand{\ppbar}{{p\bar{p}}}

\begin{document}

\title{The CDF L2 Track Trigger Upgrade} 

\author{A. Lister}
\affiliation{University of California at Davis, CA 95616, USA}

\begin{abstract}
We briefly present the eXtremely Fast Tracker stereo track upgrade for the CDF Level 2 trigger system. This upgrade enables full 3D track reconstruction at Level 2 of the 3-Level CDF online triggering system. Using information provided by the stereo layers of the Central Outer Tracker, we can decrease the fake rate by requiring the tracks to be consistent with a single vertex in all three dimensions but also by using the track information to ``point'' to the various detector components.
We will also discuss the effectiveness of the Level 2 stereo track algorithm at achieving reduced trigger rates with high efficiencies during high luminosity running.
\end{abstract}

\maketitle

\thispagestyle{fancy}

\section{INTRODUCTION} 
Since 2001 the CDF II detector has been collecting data used to carry out a rich physics program at the Fermilab Tevatron $\ppbar$ collider. The quality and purity of the data relies heavily on the high performance and efficiency of the CDF trigger system. The online trigger most heavily relied on for physics analyses is the eXtremely Fast Tracker (XFT), a track trigger system used to identify charged tracks. Tracks are identified based on the data from sense wires in the central tracking chamber (COT) which are read out in time for the Level 1 trigger decision. The tracks are extrapolated to the calorimeter and muon chambers to form electron and muon trigger candidates. 

The steady increase of the Tevatron instantaneous luminosity results in higher detector occupancy from multiple $\ppbar$ interactions. For the track trigger, this produces overlapping patterns of hits which can be identified as high momentum track, leading to the rapid growth of track trigger rates with instantaneous luminosity. The control of trigger rates by either raising the thresholds or prescaling the triggers results in signficicant loss in acceptance for important physics signatures with a negative impact on the physics program.  An upgrade of the original system was thus required. Without the track trigger upgrades the physics potential of CDF would have be significantly compromised.

A description of the axial XFT system and the Level 1 track trigger upgrade can be found elsewhere~\cite{L1}. Here we describe the Level 2 part of the XFT system. The rough sequence of the data is that track segments identified at Level 1 are delivered, with a higher resolution, to the Level 2 processor, where more available time allows for full 3D-track reconstruction. This improves not only the track trigger purity but also allows the use of the track $\phi$ and cot $\theta$ information.

\section{LEVEL 2 STEREO UPGRADE}
The Level 2 stereo upgrade is an extension of the Level 1 track trigger upgrade. It introduces a new data stream to the Level 2 trigger processors and thus significantly enhances the triggering capabilities at CDF. This data stream consists of the stereo track segements list available at Level 1 but with higher $\phi$-pixel and track slope granularity.

Because of the higher granularity of the system, and thus the increased data volume, a sparsification scheme was set up in the Stereo Finder boards~\cite{L1}. Each Level 1 $\phi$-pixel corresponds to a single bit in the Level 2 data stream mask section. If the bit is set, the full resolution information (15 bits: 3 $\phi$-pixels with 5 slope masks each) is retrieved from the chip's RAM and transfered after the mask section. The sparsified data from the two Finder FPGAs is transmitted to a Level 2 Pulsar FPGA, located on the same board, that operates at a different clock. This FPGA reformats the data and transfers is to the PULSAR boards (stands for ``PULSer and Recorder'')~\cite{PULSAR}, a general purpose board that was designed primarily as an upgrade path for the new CDF II Level 2 trigger system commissioned in the summer of 2005. For the XFT Level 2 stereo upgrade, the PULSAR baord performs the merging of the segment list data from 12 Stereo Finders and re-formats it into an S-LINK 32-bit word standard packet~\cite{SLINK}.

The S-LINK data format allows communication between the PULSAE board and an S-LINK to PCI interface, Four Input Links for Atlas Readout (FILAR)~\cite{FILAR}, card. The FILAR is a high bandwidth S-LINK to PCI interface card developed at CERN and designed to have low PCI bus utilization with minimal host processor control. Each FILAR card performs an autonomous data reception from four S-LINK channels delivering data from various detector subsystems to the memory of the Level 2 decision node PC. The PC unpacks the received data packets from the various inputs and runs the trigger algorithms. The decision packet is then sent out, through a transmitter analogous to the FILAR but with a single channel to another PULSAR board that communicates the decision to the Trigger Supervisor.

In contray to the data received from most other detector subsystems that consist of already processed information, the XFT stereo segments data is ``raw'', and before running the trigger algorithms it needs to be processed. The XFT stereo segment data volume is therefore larger than data packet from the other detector components. It ranges from 1.5 to 3 kB per event, which accounts for more than 50\% of the total data volume transferred through the PCI bus. To ensure that this enlarged data volume does not cause problems to the Level 2 PC, the XFT PULSAR boards have had abort and truncation functionality implemented. The abort functionality reduces the average demand by not transmitting any XFT stereo data on events where there are no relevant tracks and the XFT stereo algorithms would not be called. The truncation functionality terminates stereo data transmission after a predetermined number of data words, cutting off any events with unusually high data volumes and automatically accepting them if they pass all other trigger requirements.

\subsection{Stereo Track Reconstruction}
Because of the high volume of the XFT stereo data packet and the highly sparsified format, it is inefficient to unpack it fully, and therefore the unpacking is performed on demand only for regions of interest, i.e. near the position of a traversing axial track already stereo-confirmed by Level 1. The stereo track reconstruction algorithim consists of the following steps. First, the axial track is extrapolated to each of the outer three stereo superlayers. Next, at each stereo superlayer the track segments corresponding to $\pm$3 cells centered near the extrapolated $\phi$ position of the track are unpacked. Due to the $\pm 2^O$ stereo angle, pixels corresponding to the real track are displaced accordingly. The correlation between displacements at various superlayers is exploited in the Level 1 trigger, while at Level 2 the higher pixel granularity and additional time also allows the determination of the stereo properties of the track that can be expressed in terms of the distance along the beam axis between the origin of the track and the centre of the detector, $z_0$, and angle of the track relative to the beam axis, cot $\theta$,.

In the next stage, the slopes of the extrapolated track are obtained at each stereo superlayer, and the pixels with slopes inconsistent with the ones of the track are ignored. Then pixels corresponding to the same $\phi$ position are 'OR'ed between slopes. This operation decreases the amount of bit information by a factor of 5 by filtering a large number of fake track segments.

Next, the pixels fired across the three different stereo layers are combined into triplets. To decrease the combinatorics for cases where two or three adjacent pixels were fired, the mean of the cluster of pixels is used in the triplet. Only triplets that extrapolate to the luminous region of the detector ($|z_0| \leq$ 60 cm) are considered. The various triggers that make use of the stereo reconstruction individually determine which of these surviving triplet combinations is examined.

The procedure described above provides measurements of the track $z_0$ and cot $\theta$ at Level 2 with resolutions of 11 cm and 0.13 rad respectively. It is important to note that the resolution distributions are close to gaussian which ensures a high efficiency for stereo track matching to the various outer detector components.

All of the steps of the stereo track reconstruction algorithim are optimized and reduced to a sequence of bit-wise operations. Nevertheless, it is a rather CPU intensive process. To reduce the effect on the Level 2 latency of the system, stereo tracking is performed only on demand, i.e. only if the track passes all other trigger requirements.

\subsection{Trigger Rate Reduction}
The CDF II detector is operated at high instantaneous luminosities with the goal of maximizing trigger acceptance of high $p_T$ physics processes. At peak luminosity, the primary limitation of data flow is the Level 2 bandwidth of 900 Hz.

\begin{figure}
  \begin{center}
\includegraphics[width=11cm]{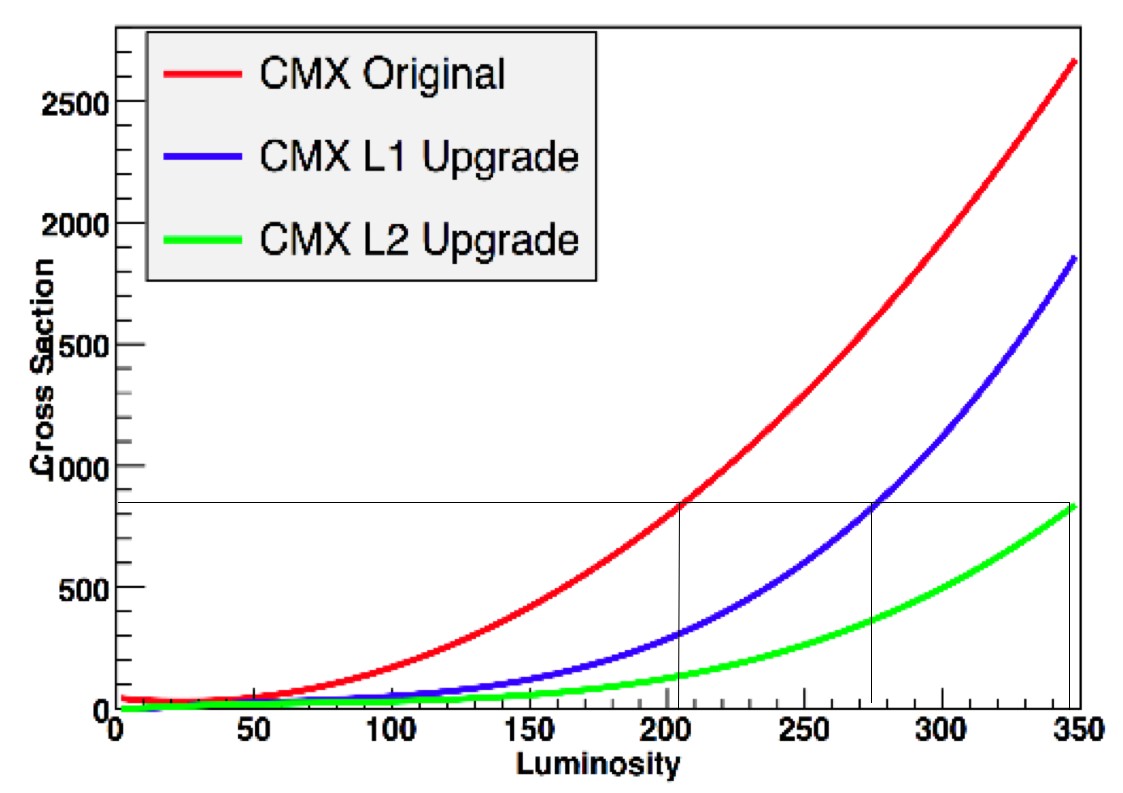}
  \end{center}
  \caption{Trigger cross sections as a function of instantaneous luminosity for the Level 2 CMX muon trigger showing the old axial layer only trigger (red) along with the Level 1 (blue) and Level 2 (green) stereo upgraded triggers.}
  \label{fig:performance}
\end{figure}

The Level 1 track trigger upgrade improves the fake track rejection factor by a factor of 3 to 5 and is ~97\% efficient. The Level 2 track trigger upgrade further reduces the trigger rate by a factor of ~3 and is over 99\% efficient. Taking as an example the high $p_T$ single muon CMX trigger, this translates to a total rate reduction, at an instantaneous luminosity of 300e30 $cm^{-2} s^{-1}$, of 2 at Level 1 and a further factor of 2 at Level 2. The CMX trigger rate is shown in Figure~\ref{fig:performance} as a function of instantaneous luminosity.

The Level 1 upgrade was completed in Fall 2006. The Level 2 part of the upgrade was comissioned in December 2007 and has been used for collecting data since then.

\section{CONCLUSIONS}

The Level 2 3D-track trigger upgrade of the CDF II detector has been described. The upgrade introduces new capabilities of 3D track reconstruction, considerably improving the performance and efficiency of the CDF II trigger system especially at high instantaneous Tevatron luminosities. It enhances the physics potential of the CDF Run II detector.

\begin{thebibliography}{9}
\bibitem{L1} 
A. Abulencia {\em et al.}, IEEE Trans. Nucl. Sci. 55, 126-132 (2008). 
\bibitem{PULSAR}
K. Anikeev, {\em et al.}, FERMILAB-PUB-06-400-E (2006).
\bibitem{SLINK}
E. van der Bij, {\em et al.}, Presented at 10th IEEE Real Time Conference. http://hsi.web.cern.ch/HSI/s-link. (1997).
\bibitem{FILAR}
W.Iwanski, {\em et al.},  Presented at 12th IEEE-NPSS Real Time Conference. http://hsi.web.cern.ch/HSI/s-link/devices/filar  (2001).
\end{thebibliography}
\end{document}